\documentclass{article}

\usepackage{PRIMEarxiv}

\usepackage[utf8]{inputenc} 
\usepackage[T1]{fontenc}    
\usepackage{hyperref}       
\usepackage{url}            
\usepackage{booktabs}       
\usepackage{amsfonts}       
\usepackage{nicefrac}       
\usepackage{microtype}      
\usepackage{lipsum}
\usepackage{fancyhdr}       
\usepackage{graphicx}       
\graphicspath{{media/}}     

\usepackage{bm}

\newtheorem{theorem}{Theorem}[section]

\newcommand{\V}[1]{
	{\boldsymbol{#1}}
}

\newcommand{\dd}[2]{%
	\frac{\mathrm d #1}{\mathrm d #2}
}

\newcommand{\eqnref}[1]{(\ref{#1})}

\usepackage{tikz}
\usepackage{pgfplots}
\usepackage{pgfplotstable}
\pgfplotsset{
compat=newest, 
tick label style={font=\footnotesize}, 
}

\title{Helmholtz's decomposition for Aeroacoustics using a standard Flow Solver
}

\author{
	\href{https://orcid.org/0000-0002-2148-6703}{\includegraphics[scale=0.06]{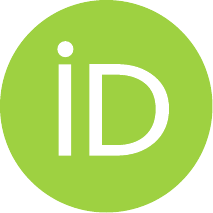}\hspace{1mm}Stefan Schoder} \\
    Institute of Fundamentals and Theory in Electrical Engineering (IGTE)\\
	Graz University of Technology\\
	8010 Graz, Austria \\
	\texttt{stefan.schoder@tugraz.at} \\
}

\begin{document}
\maketitle

\begin{abstract}
This paper is a short guideline to the decomposition of a compressible velocity into vortical and compressible structures using standard flow solvers. In particular, this is a fast solution to get an idea of the compressible fields inside your simulation, respectively acoustics for low Mach number isothermal flows. The details of the implementation are presented and the algorithm is applied to an overflown cavity with a lip. The results of the flow solver show clearly visible to acoustic radiation for the application. Finally and as previous studies showed, the use of the scalar potential formulation of Helmholtz's decomposition is valid for convex propagation domains.
\end{abstract}


\section{Introduction}
\label{sec:Intro} 
Since the beginning of computational aeroacoustics \cite{schoder2019hybrid}, the interaction effects and distinct properties of flow and sound engage interests \cite{chu1958}. Inherently, the physics of flow and sound is described by the general compressible flow equations. Therefore, aeroacoustic modeling can treat flow and acoustics directly as a united field by solving the compressible flow equations. The applications of direct aeroacoustic simulations reach from airframe noise \cite{Fares2016}, turbo-fan noise \cite{kholodov2020identification}, to acoustic simulations of landing gears \cite{Sanders2016}, and industrial applications \cite{Brionnaud}. Within this contribution, we introduce a fast version of the Helmholtz decomposition for finite domains presented in \cite{schoder2020postprocessing,schoder2018aeroacoustic,schoder2019helmholtz,schoder2020helmholtz}. The formulation can be used to a post-process a compressible velocity into vortical and compressible flow structures \cite{schoder2020helmholtz} and can effectively model the right-hand-side term of the AWE-PO \cite{schoder2023acoustic,schoder2022aeroacoustic} and cPCWE \cite{schoder2022cpcwe}.

\section{Properties of the acoustic field in isentropic flows}
According to the governing equations of fluid dynamics, acoustics is an integral part of fluid dynamics \cite{batchelor2000introduction}. The acoustic field requires a compressible fluid to be propagated. Acoustic quantities fluctuate and as a consequence, the stationary component of an acoustic quantity is zero $\overline{\V u^{\rm a}} = \V 0$. The acoustic part of the velocity field $\V u^{\rm a}$ is an irrotational field $\nabla \times \V u^{\rm a} = \V 0 $. In terms of Helmholtz's decomposition the velocity field $\V u \in \mathrm{L}^2(\Omega)$ can be separated into a vortical part $\V u^{\rm v}$ and a compressible part $\V u^{\rm a}$
\begin{equation}
\V u = \V u^{\rm v} + \V u^{\rm a}  \, .
\label{eq:acouDec}
\end{equation}
This decomposition can be accomplished for arbitrary fields. A separation into a vortical part, being a synonym for mathematically incompressible flows, and a compressible acoustic part works for low Mach numbers since all compressible effects are due to acoustics. To be more general and referring to all Mach number flows, we change the superscript from a (acoustic) to c (compressible).

\section{Formulation for isentropic flow} \label{formulation}
Based on the formulation \cite{schoder2020postprocessing,schoder2020helmholtz} that includes relevant boundary conditions and limitations, we present a method on how to integrate Helmholtz decomposition into existing CFD solvers generically. Thereby, we assume to use a convex evaluation domain. With this assumption, the scalar potential formulation of the decomposition is valid and beneficial \cite{schoder2022aeroacoustic}.
\begin{theorem}[Helmholtz decomposition for finite domains]
Every square integrable vector field $\V u \in [L_2(\Omega)]^3 $, $\mathcal{C}^1$ smooth, on a simply connected, Lipschitz domain $\Omega \subseteq \mathbb{R}^3$, has an L$^2$-orthogonal decomposition
\begin{equation}
\V u = \V u^{\rm v} + \V u^{\rm c} = \nabla \times \V A^{\rm v} + \nabla \phi^{\rm c} \, ,
\label{eq:Heq}
\end{equation}
with the vector potential $ \V A^{\rm v} \in H(\mathrm{curl},\Omega)$ and the scalar potential $\phi^{\rm c} \in H^1(\Omega)$ \cite{adams2003sobolev}. 
\end{theorem}
Regarding the theorem, the scalar potential equation (Possion's equation) reads as
\begin{equation}
\nabla \cdot \nabla \phi^{\rm c} = \nabla \cdot \V u \, ,
\label{eq:ScaLapPot}
\end{equation}
in connection with the boundaries presented in \cite{schoder2020postprocessing}. This  partial differential equation is implemented using the finite element solver openCFS \cite{schoder2022opencfs} with the correct boundary conditions accurately. This finite element implementation will not be part of the current investigation, since this paper focuses on a fast implementation using proprietary CFD solvers.

In contrast to the previously published implementation into openCFS \cite{CFS} and openCFS-Data \cite{CFSDAT}, a standard CFD simulation software solves generic transport equations and Poisson's equations. Additionally, to the computed conservation laws, the parallel computation of scalar transport equations during the CFD simulation is efficient\footnote{Since costly IO operations are mitigated.}
\begin{equation} 
\frac{\partial \rho \phi}{\partial t} + \nabla \cdot(\rho \V u \phi - \nabla \phi) = S_{\phi} \, ,
\end{equation} 
where $\phi$ denotes the scalar function and $S_{\phi} $ the source term. Neglecting the convective and unsteady term, the Helmholtz decomposition can be directly treated within the CFD simulation as scalar Poisson's equation
\begin{equation} 
- \nabla \cdot \nabla \phi = S_{\phi} \, .
\label{eq:sTEQ}
\end{equation}
The unknown source term $S_{\phi}$ is determined through a reformulation of the continuity equation 
\begin{equation}
\frac{1}{\rho}\dd{\rho}{t} +    \nabla \cdot \V u =0 \, .
\end{equation}
The Helmholtz decomposition \eqnref{eq:Heq} is inserted into the divergence of the velocity. Based on the properties of the vector potential $\V A$, the vector potential term is null and only the scalar potential $\phi$ remains in the equation
\begin{equation}
- \nabla \cdot \nabla \phi = \frac{1}{\rho}\dd{\rho}{t} = - \nabla \cdot \V u \, .
\end{equation}
By comparing the source terms of the reformulated continuity equation with the reduced scalar Poisson equation \eqnref{eq:sTEQ}, two variants of the source term are available. The first is the relative variation of the density along a streamline \cite{wu2020noise}, the second variant depends solely on the velocity
\begin{equation}
S_{\phi} = \frac{1}{\rho}\dd{\rho}{t} = - \nabla \cdot \V u \, .
\end{equation}

\section{Application for isentropic flows}
A first verification shows promising results. The decomposition using the scalar potential formulation is now applied to a flow over a cavity with a lip\cite{Henderson} in 3D, computed by a large eddy simulation of a cavity \cite{schoder2018aeroacoustic,schoder2020numerical}. 

During the post-processing of the flow in transversal and longitudinal processes, the direct sound computation of a Mach 0.15 flow past a cavity with a lip is analyzed, and it demonstrates the capabilities of the Helmholtz's decomposition for evaluating flow fields. For further details on the flow field, we refer to \cite{lazarov2018}.

Although it is an approximation of the free field \cite{schoder2019revisiting}, the free field boundary conditions are modeled by a homogeneous Dirichlet boundary. The wall boundaries are a homogeneous Neumann boundary.
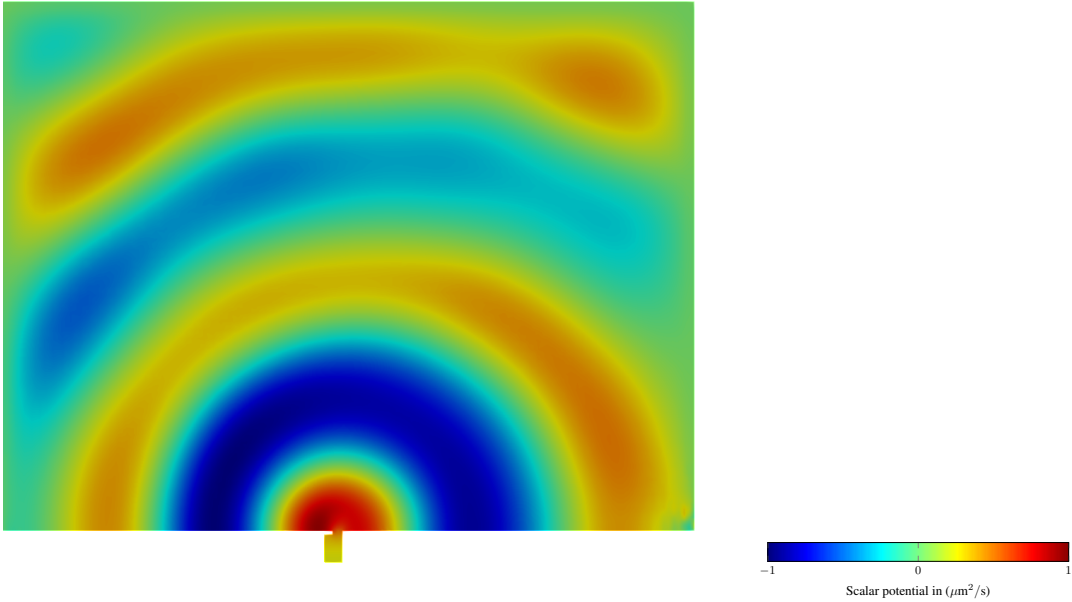
\begin{figure}[ht!] 
	\centering
		\includegraphics[width=0.6\textwidth, trim = 10cm 0cm 5cm 0cm, clip]{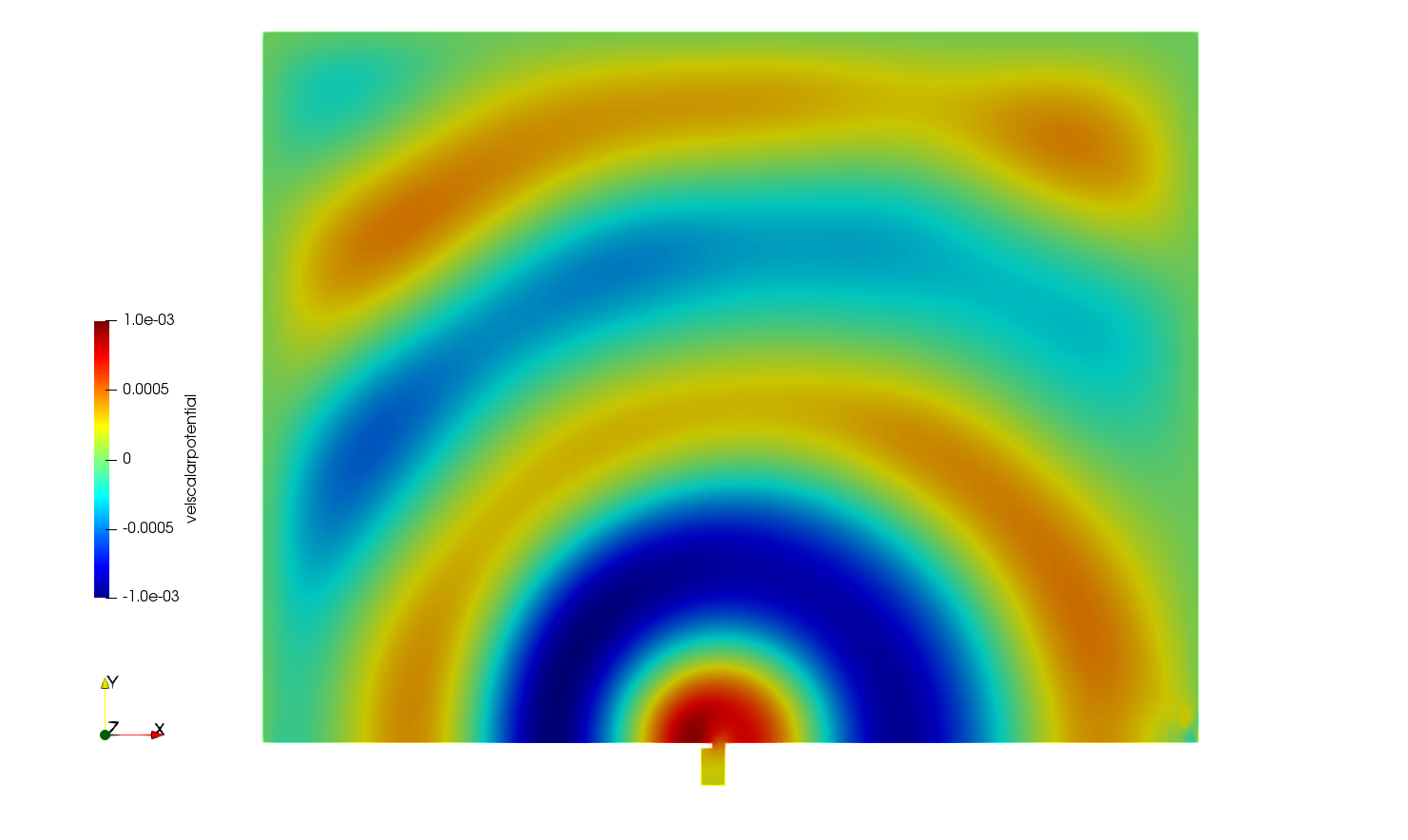}
						    \begin{tikzpicture}[scale=0.5]
\begin{axis}[
    hide axis,
    scale only axis,
    height=0pt,
    width=0pt,
    colormap/jet,
    colorbar horizontal,
    ymin=0, ymax=0.1,
    xmin=0, xmax=0.1,
    point meta min=-1,
    point meta max=1,
    colorbar style={
        width=8cm,
        xlabel=Scalar potential in ($\mu$m$^2/$s),
        xtick={-1,0,1}
    }]
    \addplot [draw=none] coordinates {(0,0)};
\end{axis}
\end{tikzpicture}
		 	 	\caption{Direct execution of the decomposition during the CFD simulation provides the scalar potential.}
 	\label{fig:decom_CFD} 
\end{figure}
Figure \ref{fig:decom_CFD} shows the result of this computation. As one can clearly see, the scalar potential extracts the compressible components of the fluid that converge to the acoustic quantities in the far-field \cite{lazarov2018}. Upstream amplification is captured by the direct simulation of flow and acoustic. The approximate free field radiation condition ''squeezes'' the acoustic field partly. The results of this formulation are available with nearly no additional simulation cost and extracts the compressible travelling sound waves. Regarding these results, the method can be used as an out-of-the-box evaluation of compressible flow simulations\footnote{Considering nearly isentropic flows.} in aeroacoustics.

\section{Properties of the acoustic momentum fluctuations and entropy momentum fluctuations}
In 1989, Doak proposed to arbitrary decompose flow fields based on the time-stationary momentum fluctuations \cite{doak1989momentum}. This time-stationary momentum fluctuations are the vector field expanded by the Helmholtz decomposition into a mean component, a vortical (turbulent), and compressible part
$$
\rho \mathbf{u}=\overline{\mathbf{B}}+\mathbf{B}^{\prime}-\nabla \psi^{\prime} \, .
$$
The compressible time-stationary momentum fluctuation is further decomposed into the acoustic component, the thermal (entropic), and species-associated components by combining the fluctuating part of the mass conservation in combination with the state equation
$$
\nabla^{2} \psi' =  \frac{\partial \rho^{\prime}}{\partial t} = \frac{\partial \rho}{\partial p} \frac{\partial p^{\prime}}{\partial t} + \frac{\partial \rho}{\partial S} \frac{\partial S^{\prime}}{\partial t} + \frac{\partial \rho}{\partial \alpha_i} \frac{\partial \alpha_i^{\prime}}{\partial t}\, .
$$ 
In this sense, the time-stationary compressible part is expressed as a linear superposition of $n+1$ potentials, depending on the modeled number of species $i \in {1,...,n}$
$$
\nabla^{2} \psi_{A}^{\prime}=\frac{1}{c^{2}} \frac{\partial p^{\prime}}{\partial t}
$$

$$
\nabla^{2} \psi_{T}^{\prime}=\frac{\partial \rho}{\partial S} \frac{\partial S^{\prime}}{\partial t}
$$

$$
\nabla^{2} \psi_{\alpha_i}^{\prime}=\frac{\partial \rho}{\partial \alpha_i} \frac{\partial \alpha_i^{\prime}}{\partial t} \, .
$$ 
where $\psi_{A}^{\prime}$, $\psi_{T}^{\prime}$, and $\psi_{\alpha_i}^{\prime}$ are the acoustic, thermal and chemical components of the irrotational part of the linear momentum density, respectively, $S$ is the entropy and $\alpha_i$ the chemical component concentration of a thermodynamically independent specie. These Poisson equations must be solved accordingly. The assumption of a time-stationary process leads to scalar compressible potential with zero temporal mean value. From these variables, the respective subfields can be recovered and used for the derivation of acoustic models for sound prediction and boundary layer modeling \cite{pierce2019acoustics}.


\section{Application}
An application of this decomposition can be found in \cite{jordan2013wave} and \cite{unnikrishnan2016acoustic} considering jet noise.

	\section{Conclusion}
	
To treat finite domains, the modified Helmholtz decomposition has been implemented and applied successfully in a proprietary finite volume solver. We showed the capabilities of Helmholtz decomposition as a post-processing tool for direct simulation of flow and acoustics for convex domains. A focus on the computational efficient scalar potential formulation yields high performance when computing the decomposition in parallel to the CFD simulation and requires no costly file-IO operations. The proposed formulation is suitable for convex domains only. The application to the convex propagation region of an overflowed cavity showed that typical radiating structures are captured. 
\section{Acknowledge}
The computational results presented have been achieved in part using the Vienna Scientific Cluster (VSC).

\bibliographystyle{elsarticle-num}
\bibliography{mybibfile}

\end{document}